\documentclass[final,aps,PRB,10pt, twocolumn, superscriptaddress, longbibliography, nobalancelastpage]{revtex4-1}
\usepackage{graphicx}
\usepackage{graphics}
\usepackage{amsmath}
\usepackage{amssymb}
\usepackage{amsfonts}
\usepackage{dsfont}
\usepackage{braket}
\usepackage[usenames,dvipsnames]{xcolor}
\usepackage{physics}
\usepackage[mathscr]{euscript}
\definecolor{darkblue}{rgb}{0, 0, 0.8}
\usepackage[colorlinks=true, breaklinks=true, linkcolor=red, citecolor=blue, urlcolor=blue]{hyperref} 
\usepackage{hyperref}
\usepackage{subfigure}
\usepackage{xfrac}
\usepackage{bm}
\usepackage{kantlipsum}
\usepackage{enumitem}
\usepackage{tikz}
\usepackage{framed}
\usepackage{booktabs}

\newcommand{\Id}{\mathds{1}}

\begin{document}

\title{Entanglement dynamics of a many-body localized system coupled to a bath} % Title

\author{Elisabeth Wybo} % Author name
\affiliation{Department of Physics, Technical University of Munich, 85748 Garching, Germany}
\affiliation{Munich Center for Quantum Science and Technology (MCQST), Schellingstr. 4, D-80799 M\"unchen}
\author{Michael Knap} % Author name
\affiliation{Department of Physics, Technical University of Munich, 85748 Garching, Germany}
\affiliation{Institute for Advanced Study, Technical University of Munich, 85748 Garching, Germany}
\affiliation{Munich Center for Quantum Science and Technology (MCQST), Schellingstr. 4, D-80799 M\"unchen}
\author{Frank Pollmann} % Author name
\affiliation{Department of Physics, Technical University of Munich, 85748 Garching, Germany}
\affiliation{Munich Center for Quantum Science and Technology (MCQST), Schellingstr. 4, D-80799 M\"unchen}

\begin{abstract}
The combination of strong disorder and interactions in closed quantum systems can lead to many-body localization (MBL). However this quantum phase is not stable when the system is coupled to a thermal environment. We investigate how MBL is destroyed in systems that are weakly coupled to a dephasive Markovian environment by focusing on their entanglement dynamics. We numerically study the third R\'{e}nyi negativity $R_3$, a recently proposed entanglement proxy based on the negativity that captures the unbounded logarithmic growth in the closed case and that can be computed efficiently with tensor networks. We also show that the decay of $R_3$ follows a stretched exponential law, similarly to the imbalance, with however a smaller stretching exponent. 
\end{abstract}

\date{\today} 
\maketitle % Insert the title, author and date

\section{Introduction}
\par Interacting quantum systems subjected to strong disorder can realize an exotic many-body localized (MBL) phase of matter~\cite{Basko2006,Vosk2013,Nandkishore2015,Abanin2019}. Similarly to a non-interacting Anderson insulator~\cite{Anderson1958}, the MBL phase is characterized by the absence of conventional transport and by spatial correlations that decay exponentially with distance. However, there are also important differences for example in the frequency-dependent response~\cite{Gopalakrishnan2015,Agarwal2017} and in the entanglement dynamics~\cite{Zinidarifmmodeheckclseci2008,Bardarson2012,Serbyn2013a}. Most prominently, the entanglement entropy grows logarithmically in time~\cite{Zinidarifmmodeheckclseci2008,Bardarson2012,Serbyn2013a}, due to effective interactions between the localized orbitals (localized integrals of motion, LIOMs)~\cite{Serbyn2013,Huse2014}. 
Evidence for an MBL phase has also been found experimentally in systems of ultracold atoms, trapped ions or superconducting qubits by the observation of a non-thermal saturation value of local densities~\cite{Schreiber842, Smith2015, Lueschen2017, Bordia2017} and by entanglement dynamics~\cite{Brydges2019, Rispoli2019, Lukin2019, Chiaro2019}. 

\par An important question is on which time scales signatures of MBL are observable in real systems, which are never truly isolated. In general, we expect that a coupling of the system to a bath leads to delocalization as transport is restored~\cite{Carmele2015,Fischer2016,Levi2016,Medvedyeva2016,Everest2017,Vakulchyk2018}. When considering Markovian dephasing noise described by the Lindblad equation, the interference between the LIOMs is destroyed and the MBL state is driven into a featureless infinite temperature state~\cite{Fischer2016,Levi2016,Medvedyeva2016}. It has been argued that local densities (e.g. the imbalance~\cite{Schreiber842}) show a universal slower than exponential (specifically, a stretched exponential) decay that can be explained in terms of the LIOMs~\cite{Fischer2016,Levi2016}. The stretched exponential decay has also been observed experimentally in a cold atom setup~\cite{Lueschen2017}. These works explain the dynamics of the imbalance in dephasive MBL systems by means of purely classical rate equations, that thus only consider the hopping of diagonal states in the density matrix. Given the recent experimental focus on entanglement dynamics in MBL systems, it is an open question of how  Markovian noise affects pure quantum correlations.

\par In this work, we investigate how the MBL phase is dynamically destroyed by a dephasive coupling to a bath, and focus on the decay of quantities that are sensitive to quantum correlations over bipartitions of the system. This is motivated by the fact that one of the most striking dynamical signatures of MBL is a generic logarithmic growth of entanglement under a quench which is completely absent in case of Anderson localization. Our goal is to investigate a recently introduced entanglement measure for open quantum systems based on the negativity, the \textit{third R\'{e}nyi negativity}~\cite{Calabrese2012,Calabrese2013,Gray2018,Wu2019}, that can dynamically capture the difference between Anderson localization and MBL provided the dissipation is sufficiently weak. We motivate our choice of this quantity, investigate how it scales, and discuss its experimental relevance. 
\par The remainder of this paper is organized as follows: We start by introducing our model and setup in Sec.~\ref{sec:model}. In Sec.~\ref{sec:open_ent} we provide some background about entanglement measures for open quantum systems, and motivate our choice to consider the third R\'{e}nyi negativity. In Sec.~\ref{sec:results} we present our computations of this quantity and its distribution over disorder realizations, and conclude in Sec.~\ref{sec:concl}.

\begin{figure}
	\centering
	\includegraphics[width = 0.9\linewidth]{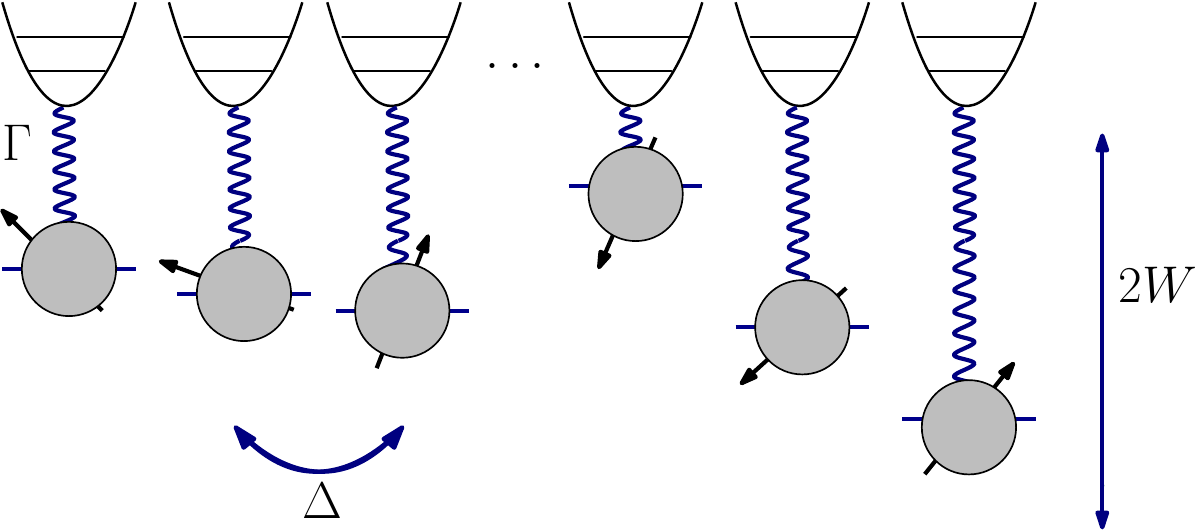}
	\caption{A sketch of our setup: a spin chain with disordered z-directed fields $h_i \in [-W,W]$, spin exchange $\Delta$ and coupling $\Gamma$ to a bath.}
	\label{fig:sketch_chain}
\end{figure}

\section{Model and setup} \label{sec:model}
\par The random-field XXZ Hamiltonian on a chain with open boundary conditions is given by
\begin{equation}\label{eq:xxz}
H =  J \left[  \sum_{i=1}^{L-1} \left( S^{x}_{i}S^{x}_{i+1} + S^{y}_{i}S^{y}_{i+1} 
+ \Delta  S^{z}_{i}S^{z}_{i+1}  \right)\right]   + \sum_{i=1}^L h_i S^{z}_{i},
\end{equation}
where $S^{x,y,z}_i$ are the spin-$\frac{1}{2}$ operators, and the $h_i$'s are randomly and uniformly distributed in the interval $[-W,W]$. Here, we will fix the disorder to $W=5J$ such that our systems are in the MBL phase~\cite{Pal2010}. 
\par The time dependence of the density matrix is given by the Lindblad master equation~\cite{Breuer2007}
\begin{equation}
\dot{\rho} = -i\left[ H, \rho(t) \right] +  \sum_i \left( L_i \rho(t) L_i^{\dagger}  -  \frac{1}{2} \left\lbrace L_i  L_i^{\dagger} , \rho(t) \right\rbrace \right) \equiv  
\mathcal{L}(\rho),
\end{equation}
which models the coupling of the system to a Markovian, i.e. memoryless, bath. We consider the Lindblad operators for dephasing noise $L_i = \sqrt{\Gamma} S_i^z $, with this choice the decoherent part of the Lindblad equation also conserves the total magnetization $\sum_i S_i^z$ in the system. The Lindblad equation then takes the form
\begin{equation}
\label{eq:lb}
\dot{\rho} = -i\left[ H, \rho(t) \right] + \Gamma \sum_i \left( 
S_i^z \rho(t) S^{z}_{i} -  \frac{1}{4} \rho(t) \right).
\end{equation}
We sketch our system in Fig.~\ref{fig:sketch_chain}. In the limit of a purely dephasive coupling $H = 0$ the off-diagonal matrix elements of the density matrix just decay exponentially. So this specific choice of Lindblad operators removes entanglement over time.
\par To simulate the time evolution accoring to the Lindblad equation, we use exact diagonalization and the time evolving block decimation (TEBD) algorithm on density matrices~\cite{Verstraete2004,Zwolak2004}, where the time-evolution operator is given by $U(t) = \exp(\mathcal{L}t)$ with the superoperator
\begin{equation} \label{eq:superop}
\mathcal{L} = -i H \otimes \Id + i \Id \otimes H + \Gamma \sum_i \left(S_i^z \otimes S^{z}_{i} -  \frac{1}{4} \Id \otimes \Id \right).
\end{equation}
\par As initial state of the quench we use the N\'{e}el product state $\rho_0 = \ket{\psi_0}\bra{\psi_0}$ with $\ket{\psi_0} = \ket{0 1 0 \dots}$, so we work in the sector with total magnetization $M = \sum_i \ev{S_i^z} =0$.
\par  The time-evolution operator acts on a vectorized version of the density matrix in which the spin indices are combined as $\ket{\rho(t+\dd t}=\exp(\mathcal{L}\dd t)\ket{\rho(t)}$. We note that the efficiency of our TEBD simulation of the density matrix critically depends on the entropy of $\ket{\rho(t)}$ viewed as a pure state in operator space. This operator-space entropy cannot be easily related to the quantum entanglement of the density matrix as it also contains classical correlations~\cite{Prosen2007}. In our setup, it has been shown that this quantity grows logarithmically which allows for a simulation over long times~\cite{Zinidarifmmodeheckclseci2008,Medvedyeva2016}. At late times the operator-space entropy converges to a value set by the steady state, which in our case is the identity restricted to the $M=0$ sector, and scales as $\log L$~\cite{Medvedyeva2016}. We provide some further numerical details in Appendix~\ref{app:numerical}.

\section{Entanglement in open quantum systems} \label{sec:open_ent} 
\par In order to motivate the entanglement quantities we are interested in, let us review some general terminology about bipartite mixed state entanglement~\cite{Guehne2008}. Consider the density matrix $\rho$ of a generic quantum system. If it is possible to decompose the density matrix as a convex sum over a bipartition 
\begin{equation} \label{eq:classical_superpos}
\rho = \sum_i p_i \rho^A_i \otimes \rho^B_i
\end{equation}
with $\sum_i p_i = 1$ and $p_i \geqslant 0$, the density matrix is called separable, meaning that there are \textit{only} classical correlations between part $A$ and $B$. If it is not possible to write such a decomposition the density matrix is quantum entangled as it is impossible to produce it locally. Now the question is to find an entanglement measure that is only sensitive to these quantum correlations. Typical entanglement monotones in the pure state formalism such as the von Neumann entanglement entropy and the R\'{e}nyi entropies 
\begin{equation}
S_q(\ket{\psi}) = \frac{1}{1-q} \log \Tr \rho_B^q
\end{equation}
with reduced density matrix $\rho_B = \Tr_A \ket{\psi}\bra{\psi}$ are not good measures in the mixed case since they are also sensitive to classical correlations.
\par It is a hard problem to determine whether or not a density matrix is separable over a bipartition as in~\eqref{eq:classical_superpos}; in fact it has been proven to be NP-hard and no general solution is known~\cite{Gurvits2003}. However there are some easily computable criteria that determine if the density matrix is not separable. They can in some cases distinguish entangled and separable states. The most important of such criteria are based on positive maps, or on entanglement witnesses~\cite{Guehne2008}. 
\par The well known criterion of separability of Peres is based on a positive map (transposition) and investigates whether the density matrix remains positive under partial transposition of a subsystem: if $\rho$ is separable $\Rightarrow$ $\rho^{T_B}$ is positive~\cite{Peres1996}. The negativity is an entanglement monotone that measures the violation of this criterion~\cite{Vidal2002} 
\begin{equation}
\mathcal{N}(\rho) = \frac{\|\rho^{T_B}\|_1-1}{2},
\end{equation}
in which the trace norm is defined as $\| O \|_1 = \Tr \sqrt{O O^{\dag}}$, hence a summation over the absolute values of the eigenvalues. The trace is invariant under partial transposition, i.e. $1 = \Tr \rho = \Tr \rho^{T_B}$, in which we assumed a normalized density matrix. Therefore
\begin{equation}
\mathcal{N}(\rho) = \sum_{\lambda_i < 0} |\lambda_i|,
\end{equation}
is effectively just a summation over the negative eigenvalues introduced in the density matrix under partial transposition. We can also define the logarithmic negativity $\mathcal{E}(\rho) = \log \|\rho^{T_B}\|_1 $. The negativity has been previously studied in the context of MBL in closed quantum systems in Ref.~\cite{Gray2019,West2018} and has been experimentally measured between two quibts in Ref.~\cite{Chiaro2019}. Note that the negativity dynamics in open quantum systems can be non-asymptotic, however in our setup we avoid this `sudden-death dynamics' by explicit spin conservation as we explain in Appendix~\ref{app:sudden_death}. The negativity is difficult to compute using tensor-network approaches. A more accessible quantity is the R\'{e}nyi negativity, much in the spirit of the R\'{e}nyi entropy in the pure-state formalism, using a replica construction
\begin{equation} \label{eq:renyi_neg}
\mathcal{E}^q(\rho) = \log \Tr (\rho^{T_B})^q. 
\end{equation}
This replica construction for entanglement negativities has been proposed in the context of field theories in Refs.~\cite{Calabrese2012,Calabrese2013,Rangamani2014}. Unlike the R\'{e}nyi entropies for pure states, the R\'{e}nyi negativites for mixed states are no entanglement monotones. However the moments of the partially transposed density matrix can be used to estimate the negativity as shown in Ref.~\cite{Gray2018}. It is also easy to see that the analytic continuations of~\eqref{eq:renyi_neg} are different for even ($q_e$) and odd ($q_o$) powers. We have that $\lim_{q_e \rightarrow 1} \mathcal{E}^{q_e}(\rho) = \mathcal{E}(\rho)$ while $ \lim_{q_o \rightarrow 1} \mathcal{E}^{q_o}(\rho) = \log \Tr \rho^{T_B} = 0$ due to the the normalization of the density matrix. For pure states, we can work out the powers of the partially transposed density matrix in terms of the reduced density matrix~\cite{Calabrese2013} 
\begin{equation}
\Tr (\rho^{T_B})^q = \begin{cases}
    \Tr \rho_B^{q_o}, & \text{$q_o$ odd,}\\
    (\Tr \rho_B^{q_e/2})^2, & \text{$q_e$ even.}
  \end{cases}
\end{equation}
By taking the limit $q_e \rightarrow 1$, we see that the logarithmic negativity can be linked to the R\'{e}nyi entropy of order $1/2$ for a pure state~\cite{Calabrese2013}
\begin{equation}
\mathcal{E}(\ket{\psi}) = 2 \log \Tr \rho_B^{1/2} = S_{1/2}.
\end{equation}
As entanglement proxy we will consider
\begin{equation}\label{eq:R3}
R_q(\rho) = -\log \left(  \dfrac{\Tr (\rho^{T_B})^q}{ \Tr\rho^q}\right) = \log \Tr \rho^q - \mathcal{E}^q(\rho),
\end{equation}
as this quantity remains zero for diagonal density matrices. For $q=1,2$ $R_q$ vanishes, such that the first non-trivial quantity is $R_3$. This R\'{e}nyi negativity gives always zero for product states, but is not necessarily zero for all separable (classically correlated) states, and hence is no entanglement monotone. We work out $R_3$ for the 2-qubit Werner state as an example in Appendix~\ref{app:r3}. It is however a computable, and potentially measurable, entanglement estimator. Note that for a pure state $\ket{\psi}$
\begin{equation}
R_3(\ket{\psi}) = -\log (\rho^{T_B})^3 = -\log \Tr \rho_B^3 = 2 S_3.
\end{equation}
\par
\begin{figure}
\centering
 \includegraphics[width = 0.99\linewidth]{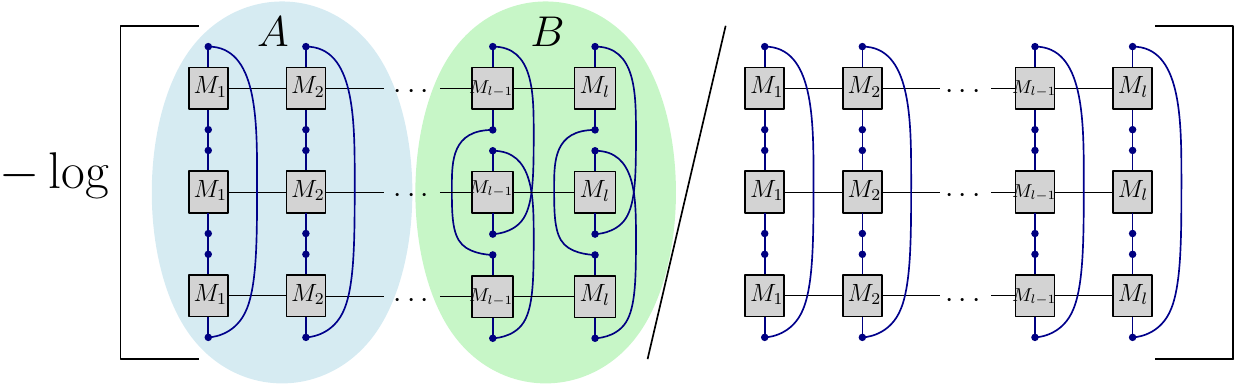}
\caption{A sketch of the $R_3$ entanglement measure, see Eq. (\ref{eq:R3}), that we compute using tensor-network techniques. We make a bipartition of the system into two subsystems $A$ and $B$, and partially transpose the degrees of freedom of subsystem $B$ before taking the trace. } 
\label{fig:sketch_r3}
\end{figure}

 $R_3$ has been previously studied in Ref.~\cite{Wu2019} in the context of finite-temperature phase transitions. In our context, the N\'{e}el state at $t=0$ and the maximally mixed state at $t=+ \infty$ have a value of $R_3 = 0$ as their density matrices do not have any negative values. At intermediate times, when there is some entanglement, the trace of the partially transposed density matrix is reduced meaning that $\frac{\tr (\rho^{T_B})^3}{\tr \rho^3}<1$, and thus $R_3 > 0$. We sketch how $R_3$ is computed using tensor-network techniques in Fig.~\ref{fig:sketch_r3}. So basically we need to compute $1.5$ full contractions of the three layer network, because the partial contraction over subsystem $A$ is used in the same way in the numerator and denominator.
\par Another possibility to quantify entanglement in open quantum systems is given by entanglement witnesses. They have the advantage that they can be experimentally relevant, because often they rely on simple expectation values. Their disadvantage is that the most optimal witness in general requires an optimization over the full Hilbert space, we discuss this further in Appendix~\ref{app:qfi}. 

\section{Results} \label{sec:results}
\subsection{Closed system}
\par
\begin{figure} 
\centering
\includegraphics[width=0.99\linewidth]{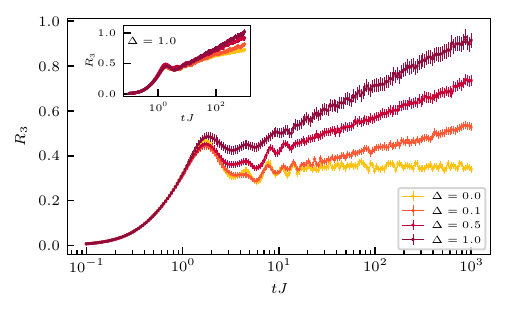}
\caption{Quench dynamics of $R_3=2 S_3 $ in the closed quantum system at fixed system size $L=14$ for various interaction strengths. In the inset we show the finite size scaling of $R_3$ at fixed interaction strength $\Delta=1$, from bottom to top $L=10,12,14,16$. The errorbars show the standard error of the mean.}
\label{fig:r3_pure}
\end{figure}

 As we have shown in the previous paragraph $R_3$ reduces to the third R\'{e}nyi entropy in the closed quantum system. 
To check that it indeed shows the same characteristic features as the von Neumann entropy in the thermodynamic limit, we have plotted its behavior in Fig.~\ref{fig:r3_pure}. We observe the typical logarithmic growth in $R_3$ for MBL systems, and a fast saturation for Anderson localized systems.
\par
\begin{figure} 
\centering
\includegraphics[width=0.99\linewidth]{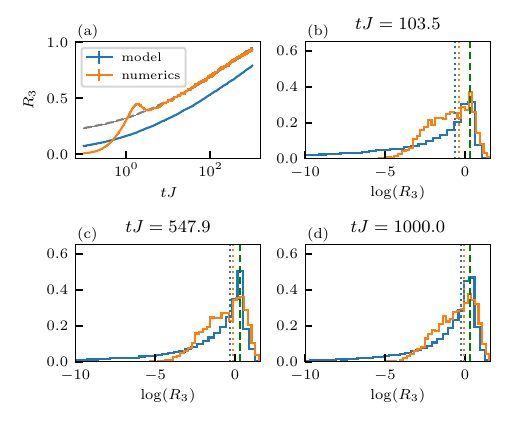}
\caption{(a) Comparison of the average of $R_3$ obtained by the numerics over $2500$ ensembles with system size $L=16$ and $\Delta =1$ and by the model described in the text with $\xi_1 = 0.40$ and $\xi_2 = 0.195$. The dashed grey line shows the curve of the model shifted by a constant. (b)-(d) Normalized histograms of the model and numerics at several times. The dotted vertical lines indicate the mean, while the dashed green line indicates the resonance at $2 \log 2$. The tails in the distributions of the model are more pronounced, as it does not take into account multi-spin couplings. }
\label{fig:r3_pure_dist}
\end{figure}

 As can be seen in Fig.~\ref{fig:r3_pure_dist}, the distributions of $R_3$ are very broad, and it is our goal to understand the shape of the distribution. Therefore we start from a simplified LIOM Hamiltonian neglecting couplings between three or more LIOMs
\begin{equation} \label{eq:tau_h}
H = \sum_i h_i \tau_i^z + \sum_{i>j}J_{ij}\tau_i^z \tau_j^z 
\end{equation}
in which the $J_{ij}$'s decay exponentially with the distance between the spins $J_{ij} = J_0 e^{-r/\xi}$ with $r = i-j$. 
\par Assuming an initial product state of two spins $\ket{\psi(0)} = \frac{1}{\sqrt{2}} ( \ket{0} + \ket{1} ) \otimes \frac{1}{\sqrt{2}} ( \ket{0} + \ket{1} )$, which are generated for the LIOMs because our initial state is prepared in a product state for the physical spins. We obtain for the entanglement generated under time-evolution with Hamiltonian~\eqref{eq:tau_h}, that  
\begin{equation} \label{eq:r3}
 R_3(t;J_r)  = -\log \left( \frac{5+3\cos(t J_r)}{8}\right), 
\end{equation}
hence the maximum $R_3$ that can be generated between the spins is $2\log 2$ as expected.
\par The couplings $J_{r}$ have been shown to be distributed according to a log-normal distribution~\cite{Varma2019}
\begin{equation} \label{eq:pJ}
P_{J}(J;r,\xi_1,\xi_2) = \sqrt{\frac{\xi_2}{8 \pi r }} \frac{1}{J} \exp\left( - \dfrac{\left( \log J  + 2r/\xi_1 \right)^2}{8 r/\xi_2} \right) .
\end{equation}
The parameters $\xi_1$ and $\xi_2$ characterize respectively the growth of the mean and variance with distance between the spins. Then we can estimate the distribution of $R_3$ for a bipartition of size $L$ by summing over $L$ values of $R_3$ that are calculated by sampling the $J$'s from $L$ distribution functions~\eqref{eq:pJ}
\begin{equation}
R_3(t) =   \sum_{r=1}^L R_3(t;J_{r}) .
\end{equation}
The average and some histograms given by this model are compared to the numerics in Fig.~\ref{fig:r3_pure_dist} for $L=16$ and $\Delta =1$. We have taken the parameters $\xi_1$ and $\xi_2$ of the distribution such that the growth of our model has the same slope ($\xi \log t$) as our data, and such that $\xi_1/\xi_2 \approx 2$ as reported in Ref.~\cite{Varma2019}, see Fig.~\ref{fig:r3_pure_dist}(a). We compare the distributions obtained by the model and by the numerics in Fig.~\ref{fig:r3_pure_dist}(b)-(c) at various times. First we note a resonance at $R_3 = 2\log 2$ both in the model as in the numerics corresponding to a singlet bound over the bipartition~\cite{Singh2016}. Secondly we note that there are two simplifications in our model (i) the difference between $\tau$-spins and physical spins and (ii) the fact that we only took into account 2-spin couplings. The first simplification is reflected in the short-time dynamics. The second simplification induces too long tails in the model towards low entanglement, as we did not take into account multi-spin couplings which can also provide significant contributions to the entanglement.  
\par The distribution of $R_3$ for Anderson localized systems would decay quickly for values higher than the singlet bond $R_3 = 2\log 2$ as entanglement cannot propagate through the system, in contrast to what we observe for the MBL system.
   
\subsection{Open system}

\subsubsection{Stretched exponential decay of $R_3$}
\par
\begin{figure}
\centering
  \includegraphics[width=.99\linewidth]{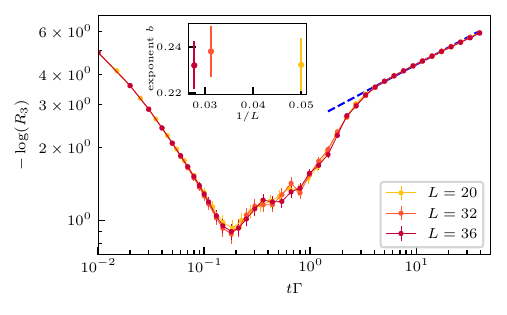}
  \includegraphics[width=.99\linewidth]{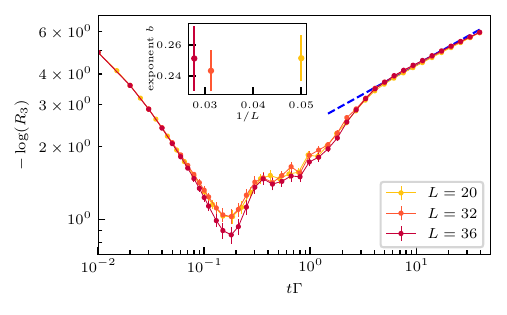}
\caption{Quench dynamics of the R\'{e}nyi negativity $R_3$ in a system with disorder $W=5J$ and coupling $\Gamma = 0.1J$. Top: MBL with interaction strength $\Delta = 1$. Bottom: Non-interacting system $\Delta=0$. In the inset we show the best fitting parameter for the stretching exponent $b$ with $3\sigma$ errorbars obtained by the least-square method. The blue line shows one of the fitting functions with $b\approx 0.25$.}
\label{fig:sc_exp}
\end{figure}

 When we switch on the dephasing, $R_3$ stays a good entanglement measure unlike the von Neumann and R\'{e}nyi entropies. In the open system, $R_3$ undergoes a characteristic stretched exponential decay starting at time scales $\gtrsim 1/\Gamma$ as shown in Fig.~\ref{fig:sc_exp}. Such a decay can be understood as a superposition of local exponential decays, and has also been observed in the imbalance %$\mathcal{I}=\frac{\expval{S_e^z-S_o^z}}{\expval{S_e^z+S_o^z}} $ 
in Refs.~\cite{Fischer2016,Levi2016,Everest2017} and is also experimentally confirmed~\cite{Lueschen2017}. We observed such a decay as well in our exact simulations for other entanglement measures like the negativity and the Fisher information as we show in Appendix~\ref{app:qfi}. 

Next, we quantitatively extract the stretching exponent $b$ of $R_3 \sim e^{-({\Gamma t/a})^b}$ by considering different system sizes. To this end, we use the TEBD algorithm on density matrices, and compute $R_3$ for the rather large coupling strength $\Gamma=0.1J$. From Fig.~\ref{fig:sc_exp} we see that the exponent is around $b\approx 0.25$, and our data shows that interactions in the Hamiltonian do not influence this exponent much. This is expected to hold true as long as interactions are small compared to the disorder in the system $\Delta < W$~\cite{Everest2017}. Ref.~\cite{Fischer2016,Levi2016} respectively reported a stretching exponent $b\approx 0.38$ and $b\approx 0.42$ for the imbalance. We observe an exponent in $R_3$ that is significantly smaller indicating that entanglement is more robust than transport under dephasing.

\subsubsection{MBL at intermediate time scales}
\begin{figure}
	\centering
	\includegraphics[width=0.99\linewidth]{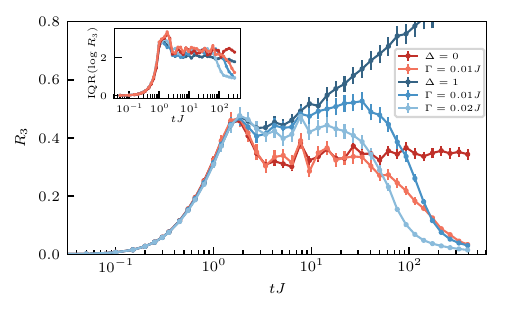}
	\centering
	\includegraphics[width=.99\linewidth]{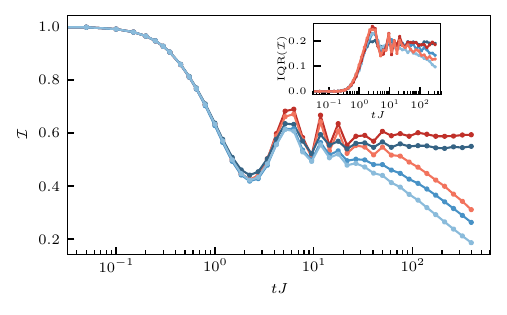}
	\caption{Dynamics of the R\'{e}nyi negativity (top) and the imbalance (bottom) under a quench in the open or closed system ($L=20$). The disorder is fixed to $W=5J$ and the interaction strength is $\Delta =1$ for the MBL system, and $\Delta =0$ for the non-interacting Anderson localized system. At intermediate time scales and for sufficiently weak dephasing strength, $R_3$ distinguishes MBL from single particle localization. A feature that is absent in the imbalance. In the insets we show the interquartile range (IQR) which forms a measure for the spread of the distribution. }
	\label{fig:open_tebd_xxz}
\end{figure}

\begin{figure} 
\centering
\includegraphics[width=0.99\linewidth]{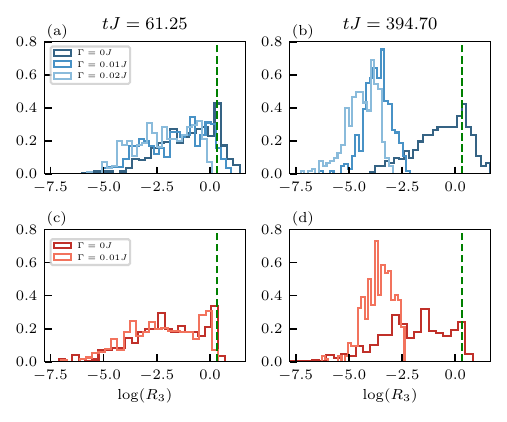}
\caption{Histograms at two different times (columns) corresponding to the averages shown in Fig.~\ref{fig:open_tebd_xxz}. (a)-(b) Interacting system with $\Delta = 1$. (c)-(d) Non-interacting system $\Delta =0$. The green line indicates $R_3 = 2\log 2$ which corresponds to the maximal entanglement between 2-spins.}
\label{fig:r3_open_dist_mbl}
\end{figure}

\par As we have seen in the previous section, interactions in the system are only a subleading effect in the stretched exponential tails, and hence these tails do not allow one to distinguish MBL from Anderson localization. The entanglement quantity $R_3$ is however able to distinguish MBL from Anderson by determining the maximally reached value of $R_3$ at intermediate time scales, on the condition that the dephasing is sufficiently weak and the interactions are sufficiently strong, i.e. $\Gamma/J < \Delta $ see Fig.~\ref{fig:open_tebd_xxz}. This is because the dephasing typically set in at a time $t \sim 1/\Gamma$ and the logarithmic growth at a time $t \sim (\Delta J)^{-1}$. We also compare the entanglement dynamics to the relaxation dynamics of the imprinted density pattern, as measured by the imbalance, $\mathcal{I}=\frac{\expval{S_e^z-S_o^z}}{\expval{S_e^z+S_o^z}} $, where $S_e^z/S_o^z$ sums $S_i^z$ over even/odd sites.  

\par We also compute the interquartile range (IQR) of our data, shown in the insets of Fig.~\ref{fig:open_tebd_xxz}, which is a measure for the spread of a distribution that is less sensitive to the tails than the variance. We choose this measure because of the limited number of ensembles that are numerically feasible for open quantum systems, which implies that we only have limited access to the tails. The dephasing is driving the state into a trivial steady state which implies that the distribution of $R_3$ will converge to a $\delta$-peak at zero entanglement. We indeed see a clear dip in the spread of the distribution as the dephasing sets in. The full distributions of $R_3$, shown in Fig.~\ref{fig:r3_open_dist_mbl}, possess strong tails even in the presence of dephasing noise, however, their width is decreasing over time. 

\par The fact that $R_3$ carries only traces from MBL at intermediate times, when the dephasing is not yet completely dominating the dynamics can also be seen from the distributions: if we want to detect traces of MBL, we need to have some larger entanglement clusters remaining over the biparition in some ensembles. Thus the distribution of $R_3$ must contain some part that has an entanglement that is higher than the singlet entanglement $R_3 = 2\log 2$, for MBL to be detectable. This criterion is more sensitive than just looking at the mean of $R_3$, since it focuses on the upper part of its full distribution.

\section{Conclusion} \label{sec:concl}
\par We have discussed a novel entanglement measure for open quantum systems in the context of many-body localization. We have seen that the third R\'{e}nyi negativity forms a promising measure to study the entanglement dynamics of an MBL system that is slightly coupled to a dephasing environment. $R_3$ can distinguish MBL from Anderson up to intermediate time scales as it reproduces the logarithmic growth of entanglement in the clean MBL system. In addition, we conclude that all quantities, entanglement and transport, decay according to a stretched exponential. However the stretching exponents are found to be smaller for the entanglement quantities, meaning that the late time entanglement dynamics is slower than for instance the dynamics of the imbalance under dephasing.  
\par The quantities $\Tr \rho^3$ and $\Tr (\rho^{T_B})^3$ are measurable without the need of full state tomography by performing joint measurements on $n=3$ copies of the state~\cite{Cai2008,Carteret2005,Mintert2005,Daley2012,Gray2018}. Alternatively, one could link $\Tr\rho^n$ and $\Tr (\rho^{T_B})^n$ to the statistical correlations of random measurements on a single copy of the state~\cite{Zhou2020}, by further developing the measurement protocols proposed for the R\'{e}nyi entanglement entropies~\cite{Enk2012, Nakata2017, Elben2018a, Vermersch2018, Brydges2019, Elben2019a}.  
\par For future work it would be interesting to investigate whether these novel protocols to measure entanglement in open quantum systems could be potentially experimentally as relevant as the protocols to measure R\'{e}nyi entropies in closed quantum systems. From the theoretical perspective it would be interesting to investigate how $R_3$ behaves under different forms of dissipation. In particular for non-hermitean types of Lindblad operators it would be interesting to investigate which signatures the entanglement structure of a non-trivial steady state contains.

\section{Acknowledgments}
We thank A.~Elben, S.~Gopalakrishnan and T.~Grover for useful discussions. Our tensor-network calculations were performed using the TeNPy Library~\cite{Hauschild2018}.
We acknowledge support from the Technical University of Munich - Institute for Advanced Study, funded by the German Excellence Initiative and the European Union FP7 under grant agreement 291763, the Deutsche Forschungsgemeinschaft (DFG, German Research Foundation) under Germanys Excellence Strategy-EXC-2111-390814868, Research Unit FOR 1807 through grants no. PO 1370/2-1, DFG TRR80 and DFG grant No. KN1254/1-1, and from the European Research Council (ERC) under the European Unions Horizon2020 research and innovation programme (grant agreements No. 771537 and 851161).
%We acknowledge support from the Technical University of Munich - Institute for Advanced Study, funded by the German Excellence Initiative, the European Union FP7 under grant agreement 291763, the Deutsche Forschungsgemeinschaft (DFG, German Research Foundation) under Germany’s Excellence Strategy-EXC-2111-390814868, Research Unit FOR 1807 through grants no. PO 1370/2-1, TRR80 and DFG grant No. KN1254/1-1, and DFG TRR80 (Project F8), and from the European Research Council (ERC) under the European Union’s Horizon2020 research and innovation programme (grant  agreements No. 771537 and 851161).

%%%%%%%%%%%%%%%%%%%%%%%%%%%%%%%%%%%%%%%%%%%%%%%%%%%%%%%%%%%%%%%%%%%%%%%%%%%%%%%%%%%%%%%%%%%%%%%%%%%%%%%%%%%%

\appendix

\section{Numerical details} \label{app:numerical}
\par The density matrix $\rho$ is represented as a matrix product operator (MPO) 
\begin{align*}
\rho &= \sum_{\substack{i_1,\dots, i_L \\ j_1,\dots, j_L }}^d M_{[1]}^{i_1,j_1} M_{[2]}^{i_2,j_2} \dots M_{[L]}^{i_L,j_L} \dyad{i_1,\dots, i_L}{j_1,\dots, j_L} \\
&= \sum_{i_1,\dots, i_L}^{d^2} M_{[1]}^{i_1} M_{[2]}^{i_2} \dots M_{[L]}^{i_L} ( \sigma_1^{i_1}  \otimes \sigma_2^{i_2} \otimes \ldots \otimes  \sigma_L^{i_L}),
\end{align*}
where $d$ is the local dimension of the Hilbert space, and where $\sigma_l^{2,\ldots,d^2}$ is a set of $d\times d$ Hermitean and traceless matrices (Gell-Mann matrices), and where $\sigma_l^1 = \Id$. In our spin-$1/2$ chain with $d=2$, these are just the Pauli matrices $\sigma_l^{2,3,4} = ( \sigma^x_l, \sigma^y_l, \sigma^z_l)$. The matrix $M_l^{i_l}$ has dimension $\chi_{l-1}\times \chi_{l}$ with $\max(\chi_{l-1},\chi_{l}) \leq \chi$, where $\chi$ is the maximal bond dimension. We can impose a canonical form on the MPO in the same way as for the MPS. We apply the TEBD algorithm on this MPO~\cite{Vidal2002,Verstraete2004,Zwolak2004,Verstraete2009}, which means that we make a Suzuki-Trotter decomposition of the time-evolution superoperator $\exp( \mathcal{L} \dd t )$ in terms of the two-site time-evolution superoperators $U_{i,i+1}(\delta t) = \exp(\mathcal{L}_{i,i+1} \delta t)$ with 
\begin{multline} \label{eq:two_site_superop}
\mathcal{L}_{i,i+1} = -i H_{i,i+1} \otimes \Id + i \Id \otimes H_{i,i+1} - \left(  \frac{\Gamma_{i} + \Gamma_{i+1} }{4} \right)  \Id \otimes \Id  \\  
+ \Gamma_i  \left(S_i^z \otimes S^{z}_{i} \right) + \Gamma_{i+1} \left(S_{i+1}^z \otimes S^{z}_{i+1} \right).
\end{multline}  
In our simulations we use the usual fourth order Trotter decomposition scheme which does in principle destroy the canonical form while performing the updates on all even or odd bonds, in addition because dissipation makes the time evolution non-unitairy. However this effect is very small for $\Gamma \lesssim J$, therefore we can still use this scheme with very good accuracies. In case the dissipation takes the form of dephasing noise, the complexity of the state can be reduced. (Note that the infinite temperature state $\sim \Id$ can be represented by a MPO with bond dimension $\chi =1$.) We truncate the singular values after acting with $U_{i,i+1}(\delta t)$ on a bond by only keeping the $\chi$ largest ones or by only keeping the ones that are larger than a certain $\epsilon_{\text{trunc}}$. Note that, unlike for the pure state case, this does not fully correspond to truncating in the entanglement of the density matrix, but rather in its complexity, or so-called operator-space entropy~\cite{Prosen2007}.
\par
\begin{figure}
\centering
  \includegraphics[width=.49\linewidth]{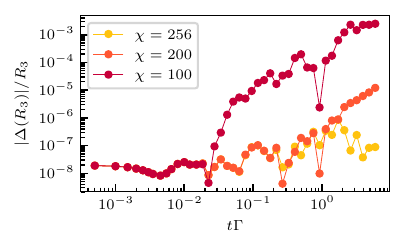}
  \includegraphics[width=.49\linewidth]{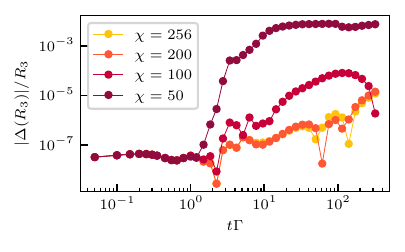}
  \includegraphics[width=.49\linewidth]{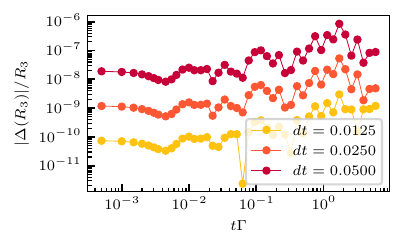}
  \includegraphics[width=.49\linewidth]{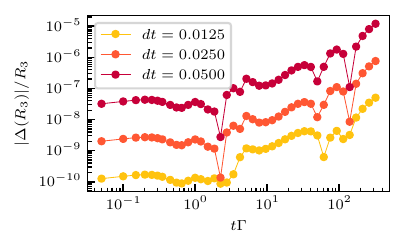}
\caption{Relative error of the fourth order TEBD scheme for various simulation parameters for a small system of $L=8$ spins. In the left column the coupling is $\Gamma = 0.01J$, in the right one $\Gamma = 1J$. In the top line we vary the maximal bond dimension. In the bottom line we vary the time step (in units of $J^{-1}$) at maximal bond dimension $\chi=256$. }
\label{fig:exact_comparison}
\end{figure}

 In the remainder of this section we show a comparison between various parameters of the TEBD on MPO algorithm. We show results for one particular disorder ensemble such that we can compare the errors caused by the algorithm, without the statistical errors from the averaging. In Fig.~\ref{fig:exact_comparison} we compare the fourth order TEBD algorithm with the exact results, by showing the relative error in the R\'{e}nyi negativity. As expected this error increases in time and with decreasing bond dimension. The main source of error that declares the small deviations from the exact result at maximal bond dimension is the Trotter error because of the splitting of the time-evolution operator. However this error can be controlled by choosing a small enough time step, as can be deduced from Fig.~\ref{fig:exact_comparison} where we also plotted the performance of the TEBD scheme at various time steps at the exact bond dimension.
\par
\begin{figure}
\centering
  \includegraphics[width=.49\linewidth]{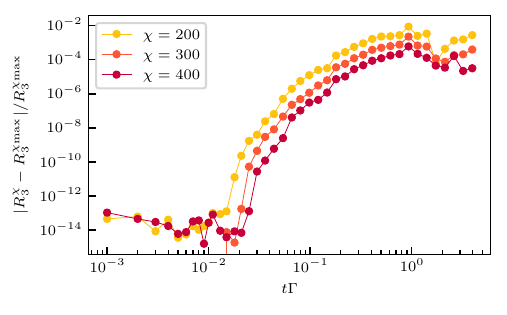}
  \includegraphics[width=.49\linewidth]{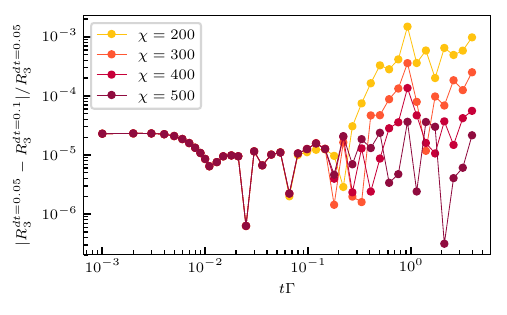}
\caption{Relative error of the fourth order TEBD scheme for $L=40$ spins with a coupling strength of $\Gamma = 0.01J$. On the left we show the relative difference between simulations with bond dimension $\chi$ and bond dimension $\chi_{\max}=500$ with a time step of $\dd t = 0.05$. On the right we show the difference between simulation with time step is $\dd t = 0.1$ and $\dd t = 0.05$ at bond dimension $\chi$.  }
\label{fig:D_comparison}
\end{figure}

 In Fig.~\ref{fig:D_comparison} we show the relative error with respect to the largest bond dimension that was easily computable for a system size of $L=40$, as well as a comparison between different time steps. From this we conclude that we maximally need a bond dimension around $\chi = 400$, and time step $\dd t =0.05$.

\section{Sudden death dynamics of the negativity} \label{app:sudden_death}
\par
\begin{figure}
\centering
  \includegraphics[width=.49\linewidth]{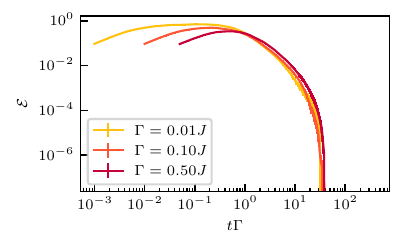}
  \includegraphics[width=.49\linewidth]{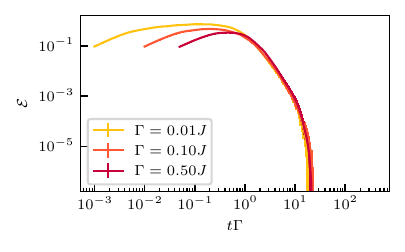}
\caption{Negativity dynamics in a small system of $L=8$ spins. When spin-conservation is broken by adding a term $ g\sum_i S_i^x$ to the Hamiltonian, the negativity dynamics stops abruptly at a finite time. In the left plot $g = 0.2J $, and in the right one $g = 0.4J$.}
\label{fig:sdd}
\end{figure}

 Entanglement quantities in open quantum systems may decay non-asymptotically, unlike transport quantities. This so-called sudden death dynamics is a known phenomenon, that imposes challenges on the stability of quantum memories~\cite{Almeida2007,Yu2009}. In our setup this specific dynamics only occurs in the negativity when we explicitly break the spin-conservation symmetry as illustrated in Fig.~\ref{fig:sdd}. In this section we investigate why the negativity decay is always asymptotic when the evolution conserves the total spin. The $U(1)$-symmetry leads to entries of the density matrix that are always zero, only one sub-block of the density matrix that corresponds to the considered spin-sector is occupied. In the $M=\sum_i \ev{S^z_i} = 0$ sector, for a chain with an even number of spins, the dimension is $m = C^L_{L/2}$.  Partial transposition maps at least part of the off-diagonal elements of the occupied sub-block to other spin sectors. Consider for instance the two qubit matrix element in the $M=0$ sector $c \ket{0 1} \bra{ 1 0}$, after partially transposing the second qubit index this becomes $c \ket{0 0} \bra{ 1 1}$, which is a matrix element outside the $M=0$ sector. Clearly under spin-conserving dynamics this matrix element would have remained zero. \\
As the diagonal elements remain of course invariant under partial transposition, we can split $\rho^{T_B}$ into two blocks $B^{out}$ and $B^{in}$, corresponding to occupied elements inside or outside the original spin sector
\begin{equation}
\rho^{T_B} = B^{out} \oplus B^{in}
\end{equation} 
with the blocks of the generic form,
\begin{equation}
B^{out} = 
\begin{pmatrix}
 & A \\
A^{\dag} & 
\end{pmatrix} \in \mathbb{C}^{2n,2n} \quad \text{and} \quad
B^{in} = (B^{in})^{\dag} \in \mathbb{C}^{m,m} 
\end{equation}
with $m+2n = \dim (\mathcal{H})$, because of the Hermiticity of the original density matrix. From this simple argument we can make no a priori assumptions about the structure of the eigenvalues of $B^{in}$. However, it is easy to see that for a density matrix of the form $B^{out}$ the eigenvalues come in pairs with opposite signs {$\pm \lambda_1, \pm \lambda_2, \dots \pm \lambda_n$}. The fact that there are always negative eigenvalues present due to inherent structure of the partially transposed density matrix of a system with spin conservation, prevents sudden death dynamics in the negativity.

\section{$R_3$ is not an entanglement monotone} \label{app:r3}
\par
\begin{figure}
 \centering
 \includegraphics[width=.99\linewidth]{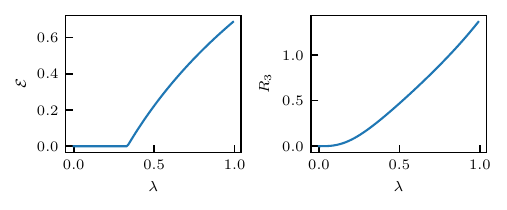}
\caption{The logarithmic negativity $\mathcal{E}$ and the third R\'{e}nyi negativity $R_3$ for the two-qubit Werner state defined by $\lambda$. $R_3$ is not an entanglement monotone as it takes a non-zero value in the separable regime $\lambda < 1/3$.}
\label{fig:werner}
\end{figure}

 By the partial transposition criterion of Peres, it follows that each separable state has a positive partial transpose. Therefore each separable state has negativity $\mathcal{E}(\rho^{\text{sep}}) = 0$, however this is not true for the R\'{e}nyi negativity. Consider for instance the two-qubit Werner state $\rho(\lambda) = \lambda \ket{\phi}\bra{\phi}+\frac{1}{4}(1-\lambda)I$ with $\lambda \in \left[ 0,1\right]$ and $\ket{\phi}$ a Bell pair, in the positive partial transpose (PPT) regime $\lambda < 1/3$. In this regime $\rho^{T_B}$ has only positive eigenvalues, and as the PPT criterion is a sufficient for separability in the two qubit case, $\rho(\lambda < 1/3)$ is separable. However the eigenvalues of $\rho^{T_B}$ and $\rho$ are not the same, which implies a non-trivial value of $R_3$. Explicitly the eigenvalues of $\rho(\lambda)$ are $\lbrace 3\times \frac{1}{4}(1-\lambda), 1 \times \frac{1}{4}(3\lambda + 1) \rbrace$, while the eigenvalues of $\rho^{T_B}(\lambda)$ are $\lbrace 3\times \frac{1}{4}(1+\lambda), 1 \times \frac{1}{4}( 1-3\lambda) \rbrace$. So in this case $R_3$ takes a non-trivial value, while $\mathcal{E}=0$. This is illustrated in Fig.~\ref{fig:werner}. Entanglement monotones satisfy invariance under LOCC (local operations and classical communication). However a separable state is transformable into any other separable by means of LOCC. (Note that local unitairy transformations fall into the class of LOCC transformations.) Therefore an entanglement monotone must remain constant over the set of separable states, which is clearly not the case for the R\'{e}nyi negativity in our example. 

\begin{figure}
\centering
  \includegraphics[width=0.75\linewidth]{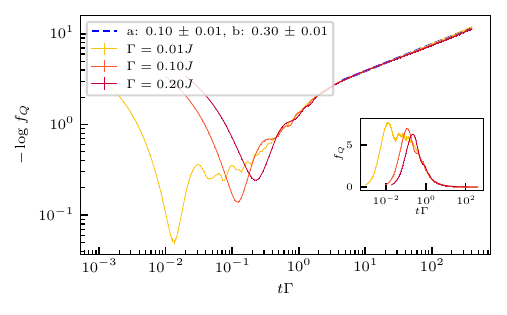}
  \includegraphics[width=0.75\linewidth]{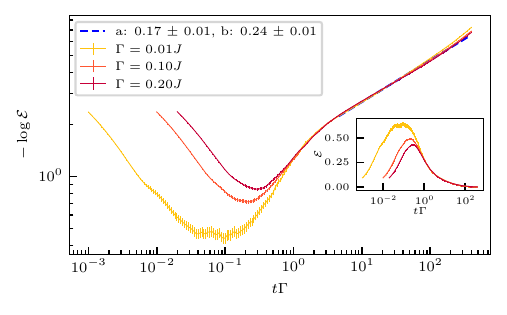}
  \includegraphics[width=0.75\linewidth]{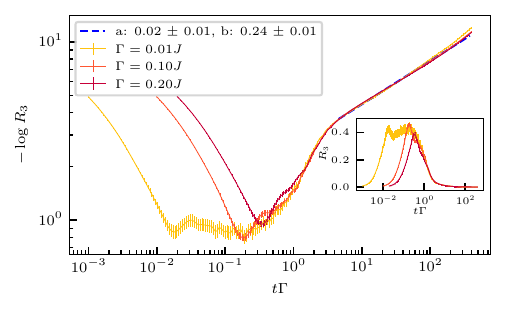}
\caption{Lindblad quench dynamics in the XXZ model at disorder $W=5J$, interaction strength $\Delta = 1$ and system size $L=8$. Stretched exponential fits $e^{-(\frac{\Gamma t}{a})^b}$ for the decay that start approximately at time $\sim \frac{1}{\Gamma}$ fit the data very well. We show from top the bottom: the Fisher information density $f_Q=F_Q/L$, the negativity and the third R\'{e}nyi negativity. The insets show the same data in a different scale. The errorbars show the standard error of the mean upon averaging over around $1000$ disorder ensembles.}
\label{fig:str_exp}
\end{figure}

\section{Quantum Fisher information as entanglement witness} \label{app:qfi}
\par Here we discuss the quantum Fisher information (QFI) which quantifies the sensitivity of a state to a unitairy transformation $e^{i\theta O}$ generated by a linear Hermitean operator of the form $O= \sum_i \bm{n}_i \cdot \bm{S}_i$, where $\bm{n}_i$ is a unit vector and $\bm{S}_i$ is the vector of spin matrices $(S^x_i, S^y_i, S^z_i)$. Therefore it measures the spread of quantum correlations via the operator $O$. The QFI witnesses entanglement in a state if its value is larger than the system size $F_Q > L$, and by other conditions it can also witness multipartite entanglement~\cite{Hyllus2012}.
For pure states, the QFI is given by the variance of $O$
\begin{equation}
F_Q(\ket{\psi},O) =  4 \left( \ev{O O}{\psi} -  \abs{\ev{O}{\psi}   } ^2 \right).
\end{equation}
For mixed states the QFI cannot be related to simple expectation values, instead the full spectral decomposition of the density matrix $\rho = \sum_i p_i \ket{s_i}\bra{s_i}$ is necessary~\cite{Braunstein1994}
\begin{equation}
F_Q( \rho, O) = 2 \sum_{\substack{i,j \\ p_i + p_j > 0 }} \frac{(p_i-p_j)^2}{p_i+p_j} \abs{\mel{s_j}{O}{s_i}}^2,
\end{equation}
and can only be computed using exact diagonalization, unless $\rho$ takes the form of a thermal state~\cite{Hauke2016}. The QFI relies on the choice of generator $O$, and for simplicity we will choose the staggered magnetization $O = \sum_i (-1)^i S_i^z$ which seems a natural choice to consider the quench dynamics from an initial N\'{e}el state. Note that the choice $O= \sum_i S_i^z $ would imply a vanishing Fisher information due to spin conservation, while $O= \sum_i S_i^x$ would imply that the Fisher information is equal to the system size for the N\'{e}el state $F_Q^{(t=0)} = L$ and under dephasing dynamics again converges to the system size $F_Q^{(t\rightarrow +\infty)} = L$. The QFI has been experimentally measured in the context of MBL in Ref.~\cite{Smith2015}. \\
\par In Fig.~\ref{fig:str_exp} we have computed the QFI, the negativity and $R_3$ by exact calculations. From this we see that the QFI also decays according to a stretched exponential, and that the stretching exponents of the negativity and $R_3$ are indeed approximately equal.

%%%%%%%%%%%%%%%%%%%%%%%%%%%%%%%%%%%%%%%%%%%%%%%%%%%%%%%%%%%%%%%%%%%%%%%%%%%%%%%%%%%%%%%%%%%%%%%%%%%%%%%%%%%%

\bibliography{biblio}

\end{document}